\documentclass[aps,nofootinbib,showpacs,showkeys,preprint
tightenlines,twocolumn
] {revtex4}

\usepackage{epsf,epsfig,subfigure,graphicx,amsmath,amssymb}
\usepackage{color}

\newcommand{\gev}{\,\textrm{GeV}}

\newcommand{\fq}{$f_{\textrm{quint}}$}
\newcommand{\mq}{$m_{\textrm{quint}}$}
\newcommand{\fqeq}{f_{\textrm{quint}}}
\newcommand{\mqeq}{m_{\textrm{quint}}}

\newcommand{\mceq}{m_{\textrm{comp}}}

\newcommand{\ac}{a_{\rm comp}}
\newcommand{\aC}{$a_{\rm comp}$}
\newcommand{\aq}{a_{\rm quint}}
\newcommand{\aG}{$\tilde\eta_h'$}

\newcommand{\three}{{\bf 3}}
\newcommand{\threeb}{{\bf\overline 3}}
\newcommand{\ten}{{\bf 10}}
\newcommand{\fiveb}{{\bf\overline 5}}
\newcommand{\five}{{\bf 5}}
\newcommand{\Nh}{{\bf N}}
\newcommand{\Nhb}{{\bf\overline N}}

\begin{document}

\preprint{SNUTP 09-003}

\title{Axionic dark energy and a composite QCD axion}

\author{Jihn E. Kim \email{jekim@ctp.snu.ac.kr}}
\affiliation{ Department of Physics and Astronomy and Center for Theoretical Physics, Seoul National University, Seoul 151-747, Korea\email{jekim@phyp.snu.ac.kr}}

\author{Hans Peter Nilles \email{nilles@th.physik.uni-bonn.de}}
\affiliation{Bethe Center for Theoretical Physics and
Physikalisches Institut,\\ Universit\"at Bonn, D-53115, Bonn, Germany
 }

\begin{abstract}
We discuss the idea that the
model-independent (MI) axion of string theory is the source
of quintessential dark energy. The scenario is completed
with a composite QCD axion from hidden sector squark
condensation that could serve as dark matter candidate.
The mechanism relies on the fact that the
hidden sector anomaly contribution to the composite
axion is much smaller than the QCD anomaly term. This
intuitively surprising scenario is based on the fact that
below the hidden sector scale $\Lambda_h$ there are many
light hidden sector quarks. Simply, by counting engineering
dimensions the hidden sector instanton potential can be
made negligible compared to the QCD anomaly term.
\end{abstract}

\pacs{12.20.Fv,14.80.Mz,95.35.+d,96.60.Vg, }

\keywords{Dark energy, Axions, Axion mixing, Hidden sector squark
condensation}

\maketitle


\section{Introduction}
Gravity mediated supersymmetry(SUSY) breaking can be realized
with a hidden sector confining force at an
intermediate scale \cite{Nilles:1982ik}.
This hidden sector and the observable sector
of the standard model couple extremely weakly through interactions
of gravitational strength.
This scheme fits very well in the heterotic string and
heterotic M-theory frameworks.
In cosmology, on the other hand, we have to deal with the dark energy (DE)
problem since more than a decade \cite{DarkEnergy}, which has triggered
a lot of interest in quintessence models \cite{quintessence88}.
In this regard, we revisit our quintessential axion
idea \cite{KimNilles03} in view of the available parameter space.

In explaining DE in terms of a quintessential axion, its vacuum
expectation value (VEV) is required not to have rolled down
until recently, otherwise the quintessential axion energy is
not DE but is cold dark matter. Of course, it is necessary for
the current vacuum energy density of the classical quintessential
axion to be of order $\lambda^4\approx (0.003~ \rm eV)^4$. These
two conditions restrict the quintessential axion decay constant
\fq~ and  mass \mq. Parametrizing the quintessential axion ($\phi$)
potential as
\begin{equation}
V[\xi]=\lambda^4 U(\xi),\quad \xi=\frac{\phi}{\fqeq},
\end{equation}
we require $\fqeq>\sqrt{(2-\delta)/6\delta}~M_P|U'|$ for $\omega=p/\rho<-1+\delta$,  where $U'=dU/d\xi$
\cite{KimNilles03}. Generically, one needs a Planckian scale
quintessential axion decay constant. So, the quintessential
axion mass is extremely small, $\lesssim 10^{-32}$ eV. Then,
there are two problems to be resolved to achieve the quintessential
axion idea: a large decay constant and extremely shallow quintessential
axion potential.

It has been believed for a long time that the model-independent (MI) axion $B_{\mu\nu}$, where $\mu$ and $\nu$ are tangential to the 4D Minkowski space,  has a rather robust model independent prediction for its decay constant \citep{Choiharm85,Svrcek06}. Recently, however, the MI axion may not be so model-independent since the decay constant depends on the compactification scheme in the warped internal space \cite{Dasgupta08},
\begin{equation}
F_a=\sqrt{\frac{2}{\beta}}~\frac{m_s^2}{M_P}
\end{equation}
where $\beta$ depends on the warping in the compact space $y\in K$,
\begin{equation}
\beta=\frac{\int d^6y \sqrt{g_{(6)}}e^{-\phi}h_w^{-2}}{ \int d^6y \sqrt{g_{(6)}}h_w^{2}},
\end{equation}
and the metric is
\begin{equation}
ds^2=h_w^2\eta_{\mu\nu} dx^\mu dx^\nu +g_{mn}(y)dy^m dy^n.
\end{equation}
Even though it is extremely difficult to find such an internal space, let us accept that there is a possibility to raise the MI axion decay constant \cite{Dasgupta08}. This makes it easier to obtain a large \fq\ from string theory.

This leads us to the question of an almost flat quintessential axion potential. Here, our main objective is to realize the QCD axion  with a decay constant at an intermediate scale and a quintessential axion with the large decay constant of the MI axion. For allocating decay constants with two confining groups, we must be careful about the axion mixing which has been discussed before \cite{Kim99,IWKim06}. In the axion mixing, there are two issues: we need to obtain two scales for the decay constants, one at the Planck scale and the other at the axion window of $10^9\sim 10^{12}$ GeV \cite{KimCarosi}, and the needed heights for the axion potentials.

The Planckian decay constant may result from the compactification as mentioned above. For the decay constant of the QCD axion in the axion window, we adopt the composite axion idea that a confining force becomes strong at the axion window  which will be somehow related to the axion decay constant \cite{Kimcomp,CompCK85}. This idea used the condensation of the `axi-color' quarks in analogy with the QCD chiral symmetry breaking. Along this line a superstring inspired composite axion was proposed before \cite{Babu94}. Here, we introduce a hidden sector for SUSY breaking and the accompanying hidden sector quarks (h-quarks). In this case, the composite axion is most probably arising from the hidden sector squark (h-squark) condensation \cite{ChunKNmu92} rather than from  an h-quark condensation due to SUSY. However, the model presented below is basically different from that of \cite{ChunKNmu92}.

The confining groups are assumed to be $SU(3)_c\times SU(N_h)$ where the hidden sector gauge group confines at $\Lambda_h\sim 10^{9-12}$ GeV. We introduce the following colored h-quarks,  transforming under $SU(3)_c\times SU(N)_h$ as
\begin{equation}
q_h=(\threeb, \Nh_h),\quad \bar q_h=(\three, \Nhb_h).\label{hquarks}
\end{equation}
We introduce $n_f$ such h-quarks so that the total number of flavors of hidden sector quarks is $N_f=3n_f$.
Due to SUSY, there also exist their superpartners, h-squarks
$\tilde q_h$ and $\tilde{\bar q}_h$. The h-gluinos are denoted as $\tilde G_h$. Superstring does not allow global symmetries except the MI axion. This means that we may not have an extra global symmetry for the QCD axion. Massless h-quarks which we will introduce are allowed only by the choice of the vacuum, not by the Lagrangian. In a vector-like SU($N$) gauge theory for example, the Lagrangian does not allow a global symmetry but a vector-like chiral field pair $Q$ and $\bar Q$ can be massless
$$
-\bar Q_LQ_R S-m^2S^*S-(S^*S)^2 -\lambda'(S)^4+{\rm h.c.}
$$
where $m^2,\lambda$, and $\lambda'$ take the values giving $\langle S\rangle=0$. In the vacuum $\langle S\rangle=0$, there exists an accidental symmetry: $Q\to e^{i\alpha\gamma_5}Q$.

The absence of the tree level mass in our scheme belongs to the so-called $\mu$ problem \cite{KimNilles84}. In some compactification schemes such as the $Z_3$-orbifold, the tree level $\mu$ term is not allowed and higher dimensional operators can lead to $\bar q_hq_h$ mass by VEVs of singlet Higgs fields \cite{CMunoz}.  In this set-up, as pointed out above some vacua may allow global symmetries by choosing vanishing singlet VEVs toward zero mass h-quarks. Thus, the h-quarks of (\ref{hquarks}) can be massless in some vacua.

\begin{figure}[!h]
\vskip 0.5cm
\resizebox{0.95\columnwidth}{!}
{\includegraphics{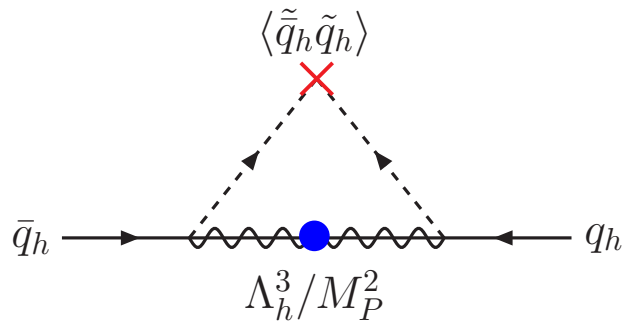}}
\caption{The one-loop h-quark mass term by the h-gluino mass ({\Large\color{blue} $\bullet$}) and chiral symmetry breaking ({\Large\color{red} {$\times$}}) insertions.}\label{fig:hqmass}
\end{figure}

Being massless, the h-squark Compton wave length is very large at the hidden sector confining scale, a pair of an h-squark and an anti-h-squark develops a correlation function at a large separation, and the h-squarks are expected to condense. Even if h-quarks are also massless, SUSY forbids an h-quark condensation.  But, if SUSY is broken, it is expected that h-quarks obtain mass also,
as a result of SUSY breakdown. The SUSY breaking in our approach here is assumed to arise from the h-gluino condensation mediated by gravitational 
interactions \cite{Nilles:1982ik}.
Then, by inserting h-gluino condensation at the intermediate scale
in the h-gluino line of Fig. \ref{fig:hqmass}, we obtain an h-quark
mass. If SUSY is broken, the h-gluino itself obtains a mass,
presumably at the TeV scale in the gravity mediation scenario.
The relevant graph for the loop induced h-quark mass is shown
in Fig. \ref{fig:hqmass}. A nonvanishing  h-quark mass requires
chiral symmetry breakdown via h-squark condensation as well as a
nonzero hidden sector gaugino mass through hidden sector gaugino
condensation at an intermediate scale.

\begin{figure}[!]
\vskip 0.5cm
\resizebox{0.95\columnwidth}{!}
{\includegraphics{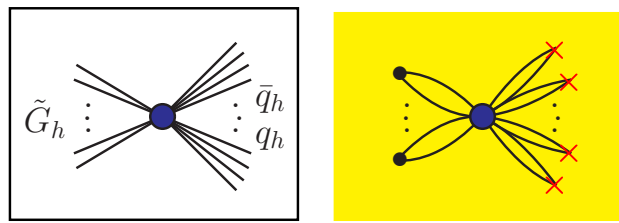}}
\caption{The 't Hooft determinental interaction shown inside the box. In the yellow shaded figure for $N_h>N_f$, the black bullet is the h-gluino condensation and the red cross is the h-squark condensation.  }\label{fig:tHooft}
\end{figure}

\section{Instanton interaction}
\begin{figure}[!]
\vskip 0.5cm
\resizebox{0.9\columnwidth}{!}
{\includegraphics{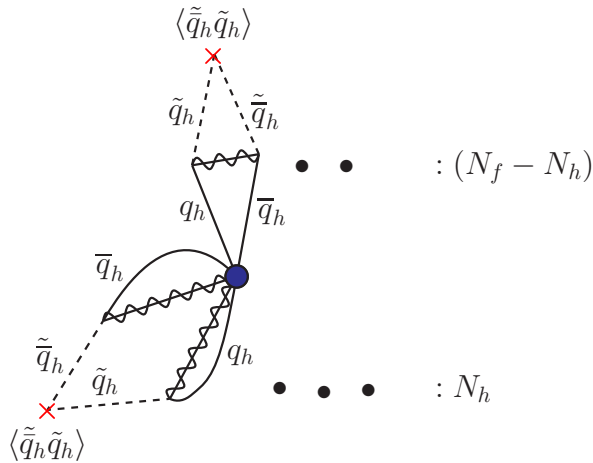}}
\caption{The instanton diagram drawn for $N_f>N_h$. The h-gluino lines are the curly ones. For $N_h>N_f$, all $2N_f$ h-squark lines are consumed by $2N_f$ h-gluino lines and there are left $(N_h-N_f)$ h-gluino condensations.  }\label{fig:tHooftNilles}
\end{figure}

Let us proceed to consider the dynamical degrees in our set up
below the condensation scale. The first diagram of Fig. \ref{fig:tHooft}
is the famous 't Hooft determinental interaction \cite{tHooft76}.
It involves 2$N_h$ gauginos as well as $N_f$ quarks and antiquarks each.
Recently, in Ref. \cite{KimCarosi} it
has been stressed that the 't Hooft determinental interaction itself
can be used for axion phenomenology. This interaction is the crucial
quantity to discuss axion masses as well as the value of the
vacuum energy for quintessential DE. In the presence of quark and
gaugino masses, we can close the lines as indicated in the
subsequent diagrams. The result has to be discussed separately
for the two cases $N_f\leq N_h$ and  $N_f>N_h$.
The h-gluino condensation produces a
$\eta'$-like particle which we will call \aG. Its mass is given
dominantly by $-m_{\tilde G_h}\tilde G_h\tilde G_h$, supplemented
by the instanton interaction diagram shown as the yellow highlighted
diagram of Fig. \ref{fig:tHooft}. The diagram considered in
Fig. \ref{fig:tHooftNilles} \cite{Nilles83} for $N_f>N_h$ has the
same order contribution. Also there appears h-squark condensation
whose phase becomes dynamical \cite{ChunKNmu92} and we will call
it another axion $\ac$. The hidden sector instanton interaction
of $N_f$ flavors of confining $SU(N_h)$ is \cite{tHooft76}
\begin{equation}
\frac{1}{K^{3(N_h+N_f)-4}}(\tilde G_h\tilde G_h)^{N_h} (\bar q_h q_h)^{N_f} \label{eq:InstInt}
\end{equation}
where $K$, roughly the inverse effective instanton size, has a
mass dimension of the hidden sector scale. Note that the
interaction (\ref{eq:InstInt}) leads to a vanishing result
in a supersymmetric theory. Axion masses and vacuum energy
will strongly depend on the breakdown mechanism of supersymmetry.
The magnitude of
(\ref{eq:InstInt}) in the broken-SUSY phase will be estimated in the next
section.

\section{Vacua of SUSY gauge theory}
The hidden-sector SUSY QCD of $SU(N_c)$ with $N_f$ flavors are reviewed 
elegantly in \cite{IntrSeiberg95}. If there is no h-gluino mass and 
h-squark mass terms, the scale entering in the effective superpotential 
is the confining scale $\Lambda\equiv \Lambda_h$. Then the low energy 
effective description of a confining theory must respect the global symmetries 
$SU(N_f)_L\times SU(N_f)_R\times U(1)_A\times U(1)_B\times U(1)_R$ 
\cite{IntrSeiberg95}, whose charges are shown in Table \ref{tab:Charges}. 
Here, $U(1)_A$ is anomalous and the remaining $U(1)$s are anomaly free.
It is made such that $U(1)_R$ is anomaly free. Considering the fermion 
zero modes in an instanton background, one assigns $2N_f/(3N_c-N_f)$ 
for the running scale $\Lambda$. In this assignment, basically we are 
considering the nonperturbative effects, contributed by the instanton 
diagrams of Figs. \ref{fig:tHooft} and \ref{fig:tHooftNilles} also. 
The $U(1)_A$ charge of $\Lambda$ given above is used mostly in the 
confining phase. Here, we comment two simple cases of \cite{IntrSeiberg95}, 
the confining phase ($N_f<N_c$) and the interacting nonabelian Coulomb 
phase ($\frac32 N_c<N_f<3N_c$).
\begin{table}
\begin{center}
\begin{tabular}{c|ccccc}
&~ $SU(N_f)_L$& $SU(N_f)_L$ &$U(1)_A$&$U(1)_B$ &$U(1)_R$\\
\hline $Q$~ & $N_f$& 1&1&1& $(N_f-N_c)/N_f$\\
 $\tilde Q$~ &1&  $\overline{N}_f$&1&--1& $(N_f-N_c)/N_f$\\
 $\tilde G$ & 1 & 1 & 0 & 0 & 1\\
 $\Lambda^{3N_c-N_f}$&1&1&$2N_f$&0&0
\end{tabular}
\end{center}
\caption{The charges of h-quark superfields $Q$ and $\tilde Q$ and of the h-gluino.} \label{tab:Charges}
\end{table}

\subsection{Case $N_f<N_c$}\label{subsec:Confine}
In this case, one obtains a dynamically generated superpotential \cite{IntrSeiberg95}, along the $D$-flat direction of $Q$ and $\tilde Q$, $Q=\tilde Q=F$,
\begin{equation}
W_{\rm dyn}=(N_c-N_f)\left( \frac{\Lambda^{3N_c-N_f}}{{\rm Det.}\tilde QQ}\right)^{1/(N_c-N_f)}. \label{eq:Seib}
\end{equation}
The dynamical term (\ref{eq:Seib}) is integrated in \cite{IntrSeiberg95} with the h-gluino condensation $S=\tilde G\tilde G$ to give the Taylor-Veneziano-Yankielowicz term \cite{VenYank82},
\begin{equation}
W(S,\Phi)=S\left[\ln\left( \frac{\Lambda^{3N_c-N_f}}{S^{N_c-N_f}{\rm Det.}\Phi}\right)+(N_c-N_f)\right]\label{eq:GlCond}
\end{equation}
where $\Phi=\tilde QQ$ and Eq. (\ref{eq:GlCond}) is restricted to be used just for estimating the h-gluino condensation scale \cite{IntrSeiberg95}. Eq. (\ref{eq:GlCond}) has the runaway solution for $\Phi$.
The confining theory with Eq. (\ref{eq:GlCond}) is the same as the theory with  Eq. (\ref{eq:Seib}) in the IR limit. In the dual magnetic phase, the coupling constant tends to zero and the gauge singlet $\Phi$ has the correct global symmetry behavior. In the magnetic phase, the confining gauge group is $SU(N_c-N_f)$ without h-quarks, and the h-gluinos in the magnetic phase would have the Veneziano-Yankielowicz effective term
\begin{equation}
W(S,\Phi)=S'\left[\ln\left( \frac{\Lambda^{'3}}{S^{'}}\right)+(N_c-N_f)\right]\label{eq:VYmag}
\end{equation}
which just determines the h-gluino condensation scale in terms of the magnetic phase parameter $\Lambda'$. The $SU(N_c-N_f)$ singlet $\Phi$ is not coupling to $S'$ and Eq. (\ref{eq:Seib}) gives the runaway solution of $\Phi$. But with SUSY breaking present at the intermediate scale, h-squark condensation scale is expected to be at the scale $\Lambda_h$.

\subsection{Case $\frac32 N_c<N_f<3N_c$}\label{subsec:IntCoul}

Another interesting case we cite here is the case of interacting non-Abelian Coulomb phase, for $\frac{3}{2}N_c< N_f<3N_c$ \cite{IntrSeiberg95},\footnote{Ref. \cite{IntrSeiberg95} lists the other cases as well.} where the elementary quarks and gluons are not confined but appear as interacting massless particles. In the dual magnetic phase, the nonabelian gauge group is with the h-color number $N_c'=N_f-N_c$ and the unchanged number of flavors, and we obtain the same region of $N_f$: $\frac32 N_c'<N_f< 3N_c'$. The VEV of an $SU(N_f)_L\times SU(N_f)_R$ invariant meson operator $\Phi=Q\tilde Q$ is parametrized as a block-diagonal matrix,
\begin{equation}
\Phi=\left(
\begin{array}{cc}
{\bf M}_{N_c\times N_c}&{\bf 0}_{N_c\times (N_f-N_c)}\\
{\bf 0}_{(N_f-N_c)\times N_c}&{\bf 0}_{(N_f-N_c)\times (N_f-N_c)}
\end{array}
\right)\label{eq:VEVPhi}
\end{equation}
where ${\bf M}_{N_c\times N_c}$ is an $N_c\times N_c$ diagonal matrix, etc. It can be seen by the value of the $N_f\times N_c$ matrix $Q$,
\begin{equation}
\Phi=\left(
\begin{array}{ccccc}
F&0&\cdots&\cdots&0\\
0&F&\cdots&\cdots&0\\
\cdots&\cdots&\cdots&\cdots&\cdots\\
0&0&\cdots\cdots\cdots&\cdots &F\\
&&{\bf 0}_{(N_f-N_c)\times N_c}&
\end{array}
\right)\label{eq:VEVQ}
\end{equation}
and $\tilde Q$ which is the transpose of the RHS of (\ref{eq:VEVQ}).
Here, there exists  a fixed point at the coupling $g=g_*$, and the theory is an interacting one. In this case, for the chiral fields $Q$ and $\tilde Q$, the gauge invariant operators $\tilde QQ$(meson) and $B$(baryon) and $\tilde B$(anti-baryon)  have the following (anomalous) dimensions and $R$ charges,
\begin{align}
&D(Q\tilde Q)=\frac32 R(\tilde QQ)=\frac{3(N_f-N_c)}{N_f}; ~ D=(1,2)\label{eq:anomdimM}\\
&D(B)=D(\tilde B)=\frac{3N_c(N_f-N_c)}{2N_f};~ D=(\frac{N_c}{2}, N_c)\label{eq:anomdimB}
\end{align}
In the electric and in the magnetic phases, the scales of the nonabelian groups are given as
\begin{equation}
\Lambda={\rm electric~ scale},\quad \tilde\Lambda={\rm magnetic~ scale},
\end{equation}
so that the relevant scale $\mu$ in the interacting nonabelian Coulomb phase is given by
\begin{equation}
(-1)^{N_f-N_c}\mu^{N_f}= \Lambda^{3N_c-N_f}\tilde\Lambda^{3(N_f-N_c)-N_f} .
\label{eq:murel}
\end{equation}
Then, the superpotential is given by \cite{IntrSeiberg95},
\begin{equation}
\frac{3}{2}N_c< N_f<3N_c:\quad W=\frac{1}{\mu}N_i^{\tilde j}(-Q^i \tilde Q_{\tilde j}+M^i_{\tilde j})\label{eq:Wmutualdual}
\end{equation}
where $M^i_{\tilde j}$ and $N_i^{\tilde j}$ are gauge singlets and  $N_i^{\tilde j}$ is the meson constructed with dual quarks. SUSY conditions fix $N=0$ and $M^i_{\tilde j}=Q^i \tilde Q_{\tilde j}$. $N$ and $M$ are massive and integrated out with the values at $N=0$ and $M^i_{\tilde j}\ne 0$ for the block-diagonal part of (\ref{eq:VEVPhi}). The anomalous dimension of $Q\tilde Q$ in the IR region is given by Eq. (\ref{eq:anomdimM}) and the anomalous dimension of $N$ is determined to give a correct dimension of $W$.

\begin{figure}[!]
\vskip 0.5cm
\resizebox{0.9\columnwidth}{!}
{\includegraphics{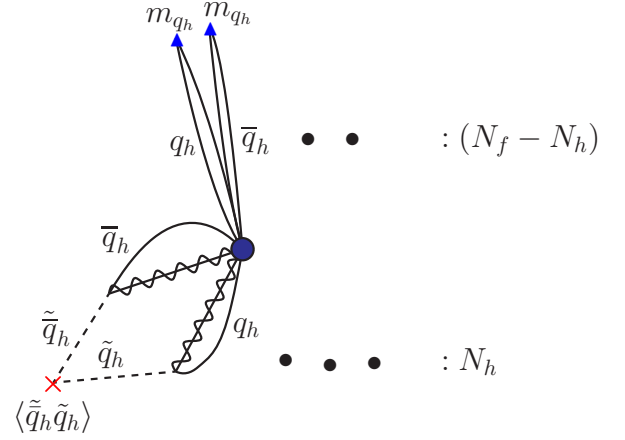}}
\caption{The 't Hooft determinental interaction drawn for $N_f>N_h$ with h-quark mass insertions (the triangles).  }\label{fig:Nilles2}
\end{figure}

After integrating out $N$ and $M$ fields, we are left with massless chiral fields $Q$ and $\tilde Q$. But the superpotential gives $\langle Q^i \tilde Q_{\tilde j}\rangle=M^i_{\tilde j}$, which means that the massless h-squarks corresponding to the block-diagonal part of (\ref{eq:VEVQ}) have a long range correlation which is like the Cooper pairs in the superconductivity. By the gauge singlet VEVs $\langle Q^i \tilde Q_{\tilde j}\rangle$ constructed with these, some global symmetries are broken. But the h-squarks corresponding to ${\bf 0}_{(N_f-N_c)\times N_c}$ of (\ref{eq:VEVQ}) do not develop VEVs. Thus, the upper part of Fig. \ref{fig:tHooftNilles} cannot be closed by the h-squark condensation. Nevertheless, we can consider the explicit chiral symmetry breaking as shown in Fig. \ref{fig:Nilles2} by the h-quark masses.

\begin{table}
\begin{center}
\begin{tabular}{c|ccccc}
&~ $SU(N_f)_L$& $SU(N_f)_L$ &$U(1)_A$&$U(1)_B$ &$U(1)_R$\\
\hline $q$~ & $\overline{N}_f$& 1&0& $N_c/(N_f-N_c)$& $N_c/N_f$\\
 $\tilde q$~ &1&  ${N}_f$&0& $-N_c/(N_f-N_c)$& $N_c/N_f$\\
 $\mu$ &1&1 &2&0&0
\end{tabular}
\end{center}
\caption{The charges of h-quark superfields $Q$ and $\tilde Q$ and of the h-gluino.} \label{tab:Chargesdual}
\end{table}

The dual magnetic theory is anomaly free as well as the electric theory \cite{IntrSeiberg95}, and hence we assign to the dual h-squarks $q$ and $\tilde q$ (also $N$) the vanishing $U(1)_A$ charge. For the scale $\mu$, we assign the $U(1)_A$ charge 2 from Eq. (\ref{eq:murel}) since $\tilde\Lambda$ is required to carry no $U(1)_A$ charge.
Then, we can integrate in $S$ as before to give
\begin{equation}
W(S,Q\tilde Q)=S\left[\ln\left( \frac{N(M-Q\tilde Q)}{\mu S}\right)+1\right]\label{eq:IninSMN}
\end{equation}


\subsection{SUSY breaking with one family of SU(5)}
\begin{figure}[!]
\vskip 0.5cm
\resizebox{0.5\columnwidth}{!}
{\includegraphics{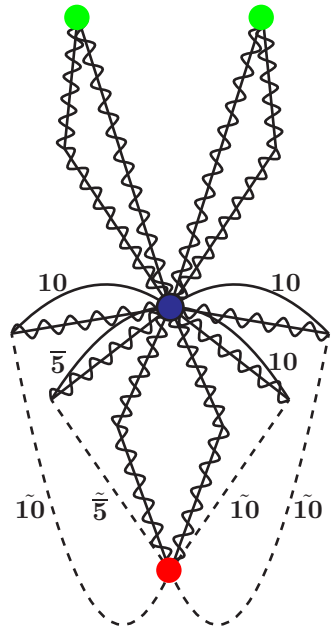}}
\caption{The $SU(5)$ instanton interaction drawn for one chiral family. The green bullets are $Z$s and the red bullet is $Z'$. }\label{fig:SUFinst}
\end{figure}

So far we introduced an extra SUSY breaking outside the $SU(N_c)$ sector. In this subsection, we discuss the one family $SU(5)$ model \cite{SUFiveVen} so that SUSY is also broken dynamically by the hidden-sector gauge group $SU(5)$. Let us consider $N_f$ pairs of $\five^i$ and $\fiveb_i$ in addition to the chiral $\ten+\fiveb$. The $U(1)$ symmetries we consider are shown in Table \ref{tab:SUfive}.
\begin{table}
\begin{center}
\begin{tabular}{c|cccc}
&~ $U(1)_A$& $U(1)_B$& $U(1)_R$ & $U(1)_C$ \\
\hline $\ten$~ & $p$&$1$& $2(N_f-10)/9$ & $0$\\
 $\fiveb$~ & $q$&$-3$& $2/3$ & $-N_f/(N_f+1)$\\
 $\five^i$~ & $q$&$3$& $2/3$ & $1$\\
 $\fiveb_i$~ & $q$&$-3$& $2/3$ & $-N_f/(N_f+1)$\\
 $Z$ &0&$0$&2 & $0$\\
 $Z'$ &$3p+q$&$0$&$2(N_f-6)/3$ & $-N_f/(N_f+1)$\\
 $\Phi$ &$2qN_f$&$0$&$4N_f/3$ & $N_f/(N_f+1)$\\
 $Z'_\Phi$ &$3p+(2q+1)N_f$&$0$&$2(N_f-2)$ & $0$\\
 $\Lambda^{3N_c-2-N_f}$ &~$3p+(2q+1)N_f$&$0$&0 & $0$
\end{tabular}
\end{center}
\caption{Here, $Z=\tilde G\tilde G, Z'=\ten\cdot\ten\cdot\ten\cdot\fiveb\cdot \tilde G\tilde G$, and $ \Phi={\rm Det.}\five^i\fiveb_j$.} \label{tab:SUfive}
\end{table}
The superpotential respecting these symmetries can be written as
\begin{equation}
W_{\rm SU(5)}=Z\left[\log\left(\frac{Z^{2-N_f} Z'_\Phi}{\Lambda^{3N_c-2-N_f}} \right) -\alpha\right]\label{eq:SUfW}
\end{equation}
where $N_c=5,~ \alpha$ is a coupling and we have not specified $p$ and $q$. For $N_f=0$, $Z'_\Phi$ reduces to $Z'$. Because of the bosonic symmetries, the combination $\ten\cdot\ten\cdot\ten\cdot\fiveb$ vanishes identically, but $Z'$ is not necessarily vanishing \cite{Meurice84},
\begin{equation}
Z'\sim \epsilon_{acfgh}{\tilde G}^{a}_b{\tilde G}^c_{d} \ten^{eb}\fiveb_e \ten^{fd}\ten^{gh}.
\end{equation}
When $N_f$ pairs of $\five^i$ and $\fiveb_i$ are added, we must consider the flavor group $SU(N_f)\times SU(N_f+1)$ also and $\Phi$ cannot appear as an independent variable, and we must consider only the combination $Z'_\phi\equiv Z'\five^1\fiveb_1\cdots \five^{N_f}\fiveb_{N_f}$ with appropriate indices contraction.
The SUSY conditions from $W_{\rm SU(5)}$ is
\begin{equation}
Z^{2-N_f}Z'_\Phi=\Lambda^{13-N_f}~e^{\alpha+N_f-2}, \quad \frac{Z}{Z'_\Phi}=0,
\end{equation}
which are mutually inconsistent for finite values of $Z'_\Phi$, except for the case of $N_f=2$. Note, in particular, that the two conditions are mutually inconsistent for $N_f=3$.  So, we obtain the dynamical breaking of SUSY for $N_f=0$ \cite{SUFiveVen} as well as for any value of $N_f$ vector-like pairs, which is consistent with the index theorem. For the $N_f=0$ case the instanton diagram is shown in Fig. \ref{fig:SUFinst}.

\section{Pseudo-Goldstone bosons}

In Subsecs. \ref{subsec:Confine} and \ref{subsec:IntCoul}, we discussed how the h-squark condensation is determined by the h-gluino condensation. From now, we go back to the original notation $N_c\to N_h, Q\to q_h, \tilde G\to  \tilde G_h$ and $\Lambda\to\Lambda_h$.  The h-gluino condensation, which preserves SUSY, is assumed to take place by some strong dynamics. These break global symmetries. In particular, we would like to deal with two Goldstone bosons, the phase of the h-gluino condensation \aG\ and the phase of the h-squark condensation $\ac$.

The kinetic energy term of a composite Goldstone boson, such as \aG\ and $\ac$, is generated when a continuous global symmetry is broken.\footnote{The axion in the usual case is not composite and the kinetic energy term is present in the Lagrangian \cite{PQ,KSVZ,DFSZ}.} For \aG\ and $\ac$, the h-gluino condensation $\langle \tilde G_h\tilde G_h\rangle$ and h-squark condensation $\langle \tilde{\bar q}_h\tilde q_h\rangle$ provide this requirement. The two global symmetries are the phase of the h-gluino, which will be called `{\it h-gluino phase symmetry}', and a $U(1)$ phase of h-squarks, which will be called `{\it h-squark U(1) symmetry}'. The mass of a pseudo-Goldstone boson is usually given by a term explicitly breaking the presumed global symmetry. So, we need two explicit symmetry breaking terms of the above global symmetries to give masses to both \aG\ and $\ac$.\footnote{In QCD, the 't Hooft determinental interaction can be considered as an explicit symmetry breaking term which is considered together with the quark mass term for the $\eta'$ mass \cite{KimCarosi}.} Here we are in the broken-SUSY phase and the 't Hooft determinental interaction discussed below is considered as a perturbation in the broken-SUSY phase. The 't Hooft determinental interaction of the hidden sector breaks one combination of the {\it h-gluino phase symmetry} and {\it h-squark U(1) symmetry}. If this 't Hooft determinental interaction is the only explicit symmetry breaking term, one Goldstone boson will remain massless. So, we need an additional explicit symmetry breaking term to give masses to both Goldstone bosons. In fact, the mass term for the h-gluino, which appears when SUSY is broken, is another explicit SUSY breaking term in addition to the 't Hooft anomaly term.

We started at a vacuum without h-quark mass. We follow the scheme where SUSY is broken by an $F$-term of a scalar multiplet (such as the dilaton) giving mass to h-gluino. Thus, the h-gluino mass is formally carrying  $R=-2$ so that $m_{\tilde G}\tilde G\tilde G$ is allowed. Because the superpotential is not renormalized with SUSY, we must insert a SUSY breaking insertion for the h-quark mass generation. Thus, in Fig. \ref{fig:hqmass} the h-gluino mass term is $V^3/M_P^2$ where $V^3$ is the h-gluino condensation scale and $M_P$ is the Planck mass. Then, with the h-squark condensation $U^2$ we can estimate the h-quark mass of order
\begin{equation}
m_q\sim \frac{g_h^2}{16\pi^2}\frac{V^3U^{*2}}{M_P^2\Lambda_h^2}
\sim 10^{-2}\epsilon^2 \rho^3 \delta^2\Lambda_h\label{mqloop}
\end{equation}
where
\begin{equation}
\epsilon= \frac{\Lambda_h}{M_P}, \quad\rho=\frac{V}{\Lambda_h},\quad \delta=\frac{U}{\Lambda_h}
.
\end{equation}
Eq. (\ref{mqloop}) carries $R=2(N_h-N_f)/N_f$ which is consistent to give mass to h-quarks of Table \ref{tab:Charges}.

\begin{figure}[!h]
\vskip 0.5cm
\resizebox{0.95\columnwidth}{!}
{\includegraphics{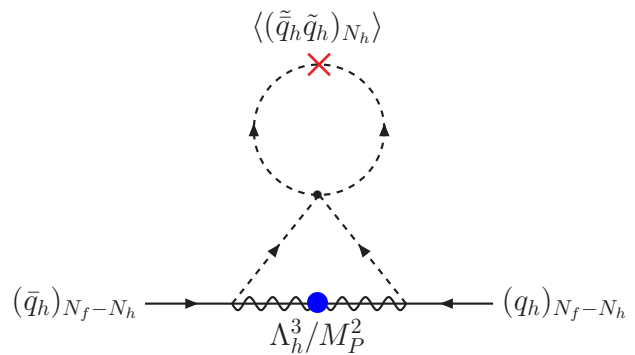}}
\caption{The two-loop $(N_f-N_h)$ h-quark mass term by the h-gluino mass ({\Large\color{blue} $\bullet$}) and chiral symmetry breaking ({\Large\color{red} {$\times$}}) insertions.}\label{fig:hqmCoul}
\end{figure}

Note, however, that in the interacting Coulomb case ($\frac32 N_h<N_f<3N_h$) of SUSY QCD, the h-squark condensation is possible only for $N_h$ h-squark pairs among the $N_f$ h-squark pairs. The h-quarks corresponding to $2(N_f-N_h)$ cannot obtain mass in this way through Fig. \ref{fig:hqmass}. We need to break the chiral symmetry of $SU(N_f-N_h)_L\times SU(N_f-N_h)_R$. This can be achieved by introducing gravitational interaction,
\begin{equation}
W([Q\bar Q]_{N_f-N_h})=\alpha_X\frac{X\bar X}{M_P}[Q\bar Q]_{N_f-N_h}\label{eq:hqmassgrav}
\end{equation}
where the lowest component of the superfield $X\bar X$ develops a 
VEV \footnote{The superfield $Q$ has $\tilde q_h$ as its lowest component.}. 
Then, the $SU(N_f-N_h)_L\times SU(N_f-N_h)_R$ symmetry is broken by the h-quark mass given by Eq. (\ref{eq:hqmassgrav}). But we started at a vacuum where the tree h-quark mass is vanishing and the VEV $\langle X\bar X\rangle$ must be vanishing. However, within our scheme we can have  $[Q\bar Q]_{N_h}$ instead of $X\bar X$, but  
$[Q\bar Q]_{N_h}[Q\bar Q]_{N_f-N_h}$ does not carry a proper $R$ charge. So, loops respecting the $R$ symmetry must be considered such as presented in Fig. \ref{fig:hqmCoul}. It has one more loop compared to Fig. \ref{fig:hqmass}. The quartic interaction in Fig. \ref{fig:hqmCoul} arises from the $SU(N_f)$ singlet component $N$ in Eq. (\ref{eq:Wmutualdual}).
For the VEV of $[Q\bar Q]_{N_h}$, the $(N_f-N_h)$ h-quark masses are
\begin{equation}
m_{N_f-N_h}\sim \left(\frac{g_h^2}{16\pi^2}\right)^2\frac{V^3U^{*2}}{M_P^2\Lambda_h^2}
\sim 10^{-4}\epsilon^2 \rho^3 \delta^2\Lambda_h\label{eq:hqmassnum}
\end{equation}

Thus, the h-instanton contribution to the $\tilde\eta_h'$ mass, arising from the yellow highlighted diagran of Fig. \ref{fig:tHooft}, is negligible compared to the mass contribution from the h-gluino mass term $-m_{\tilde G_h}\tilde G_h\tilde G_h$, unlike the $\eta'$ mass in the QCD chiral symmetry breaking case where the quark mass term is negligible compared to the anomaly term for the $\eta'$ mass. Still this yellow highlighted diagram contributes to the $\tilde\eta_h'$ mass and we will consider it in diagonalizing the pseudo-Goldstone boson masses.

The h-quark mass term $W=-m_q \bar q_h q_h$,\footnote{We obtained $m_{q}$ with the SUSY breaking insertion and $W$ means just the mass term of h-quark.} if $m_q$ is a field-independent parameter, breaks the {\it h-squark U(1) symmetry}, viz. $m_{3/2}m_q\tilde{\bar q}_h\tilde{q}_h$. However, our derived $m_q$ of  Fig. \ref{fig:hqmass} is field dependent ($\tilde{\bar q}_h\tilde{q}_h$) and $W$ is read from
\begin{equation}
\int d^4\vartheta~ \frac{1}{\Lambda_h^2}
q_h\bar q_h q_h^*\bar q_h^*.\label{effqhterm}
\end{equation}
Comparing Eq. (\ref{effqhterm}) with Eq. (\ref{mqloop}), $\langle (q_h^*\bar q_h^*)_F\rangle$ is interpreted as $\sim (\alpha_h/4\pi)V^3 U^{*2}/M_P^2$. The nonrenormalizable interaction (\ref{effqhterm}) results from the strong hidden sector dynamics.
Even though the VEV $\langle (q_h^*\bar q_h^*)_F\rangle$ gives a mass for the h-quark, this h-quark mass term cannot explicitly break the {\it h-squark U(1) symmetry} as shown in Eq. (\ref{effqhterm}) and hence cannot contribute to the composite axion mass. Namely, the term $ m_{3/2}m_q \tilde{\bar q}_h\tilde{q}_h$ with $m_q$ carrying the PQ quantum number of $q_h^*\bar q_h^*$ cannot contribute to the composite axion mass.

The dominant contribution from the hidden sector instanton is the yellow-shaded diagram of Fig. \ref{fig:tHooft} and Fig. \ref{fig:tHooftNilles} which are much smaller than the h-gluino mass term $m_{\tilde G}\langle\tilde G\tilde G\rangle$. Note that it is contrary to the QCD case where the instanton contribution is the dominant one for the $\eta'$ mass. Thus, the pseudo-Goldstone boson \aG\ obtains the mass square of order $m_{\tilde G}\Lambda_h$ from the h-gluino mass term $m_{\tilde G}\langle\tilde G\tilde G\rangle$. The pseudo-Goldstone boson corresponding to $\langle \tilde q_h \tilde{\bar q}_h\rangle$ obtains mass from the yellow-shaded diagram of Fig. \ref{fig:tHooft} and Fig. \ref{fig:tHooftNilles}. The dominant terms contributing to the pseudo-Goldstone boson masses
of $\langle\tilde G_h\tilde G_h\rangle $ and $\langle \tilde{\bar q}_h \tilde q_h\rangle$ are, for $\frac32 N_c<N_f<3N_c$,
\begin{equation}
m_{\tilde G}\langle\tilde G_h\tilde G_h\rangle +\frac{1}{K^{N_f+N_h-4}}m_{q_h}^{N_f-N_h} \langle \tilde{\bar q}_h \tilde q_h\rangle^{N_h}
\end{equation}
where $K$ has a mass dimension of the hidden sector confining scale. Let us consider the gluino condensation scale differently from the hidden sector scale, $V\ne \Lambda_h$. Then, in view of (\ref{eq:hqmassnum}), the leading diagrams set the mass scale of the composite axion at through the contribution to the
potential energy of the form
\begin{equation}
\frac{\Lambda_h^{N_f+N_h}}{K^{N_f+N_h-4}}
\left(\frac{\epsilon\rho}{100}\right)^{2(N_f-N_h)}\rho^{(N_f-N_h)}
\delta^{2N_f}
\label{Vquintht}
\end{equation}
for $N_f>N_h$. To get an idea, let us take $K\approx\Lambda_h\approx 10^{12}\gev$,  $\rho\approx 10^{-1}$,  and $\delta\approx 10^{-1}$. For $N_h=5$ and $N_f=9$ (three pairs of $q_h$ and $\bar q_h$) corresponding to the interacting nonabelian Coulomb phase, Subsec. \ref{subsec:IntCoul}, we obtain $10^{-46}\gev ^4$ for Eq. (\ref{Vquintht}). Eq. (\ref{Vquintht}) has a very strong dependence on these parameters and we conclude that the quintessential axion potential can be made sufficiently flat, achieving the height of order $(\textrm{0.003 eV})^4$ for a reasonable range of parameters.

\section{Masses of axions}
In the present work, we assumed the vanishing tree level hidden sector quark masses by choosing an appropriate vacuum. Namely, the chiral symmetry is broken by nonrenormalizable interactions of the form $ \bar q_hq_h X_1X_2\cdots$ where $X_i$ are gauge singlet chiral fields. If $\langle X_1X_2\cdots\rangle=0$ is chosen by the vacuum, then our following estimate on the composite axion property is still valid without a chiral symmetry at the Lagrangian level.  In this way, we may obtain an approximate global symmetry $U(1)$ from string theory \cite{CKKapprox}  toward the composite axion. The h-quark mass from Eq. (\ref{eq:hqmassnum}), for $\rho=10^{-1}, \delta=10^{-1}$ and $\Lambda_h\approx 10^{12}$ GeV, is about $1$ eV. For the composite axion mass, we need the explicit symmetry breaking terms of the {\it `h-squark U(1) symmetry'}. These are provided by the anomaly term and the h-(s)quark mass term. {\it However, the h-quark mass we commented above is the result of spontaneous breaking of the `h-squark' U(1) symmetry and hence does not contribute to the composite axion mass directly,} as we discussed before.

We begin with three pseudo-Goldstone, \aG, \aC\ and the MI axion.
Then, integrating out the pseudo-Goldstone boson \aG, we are left with the interesting MI axion and the composite axion. These pseudo-Goldstone bosons have the explicit symmetry breaking terms through the anomalies of the hidden and QCD sectors.

\begin{figure}[!]
\vskip 0.5cm
\resizebox{0.95\columnwidth}{!}
{\includegraphics{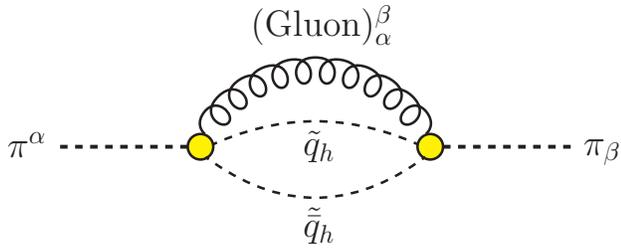}}
\caption{The h-squark lines are dashed, and the gluon line is spring-shaped. Here, $\alpha$ and $\beta$ are the color indices such as those of an octet.}\label{fig:coloredGoldstone}
\end{figure}

Let us now look for possible unwanted light Goldstone bosons lurking below the confinement scale $\Lambda_h$. The $N_f$ flavors of $U(N_h)$ h-squarks make up $N_f^2$ Goldstone bosons, and we discussed just one of them, $a_{\rm comp}$, above. The remaining $SU(N_h)$ h-squark Goldstone bosons, whose number is $N_f^2-1$, are mostly colored except two color singlet Goldstone bosons. For $N_f=9$, there are 72 colored light Goldstone bosons and 9 color singlet Goldstone bosons. For $N_f=6$, there are 32 colored light Goldstone bosons and 4 color singlet Goldstone bosons. The colored Goldstone bosons obtain mass by color interactions as shown in Fig. \ref{fig:coloredGoldstone}.  The color singlet neutral Goldstone bosons do not obtain mass by this kind of gauge interactions, and are not ruled out phenomenologically for the favoured choice $F_a\gtrsim 10^9$ GeV.

For an explicit illustration of the color singlet Goldstone bosons, let us consider the $N_f=6$ case, with the hidden-flavor symmetry $SU(2)_f$. Let the color index be $\alpha=r,g,y$ and the flavor index be $i=1,2$. The color-flavor representation of mesons is
\begin{equation}
M=\left(
 \begin{array}{cccccc}
 r1\overline{r1} & r1\overline{g1} & r1\overline{y1} &
 r1\overline{r2} & r1\overline{g2} & r1\overline{y2} \\ \\
 g1\overline{r1} & g1\overline{g1} & g1\overline{y1} &
 g1\overline{r2} & g1\overline{g2} & g1\overline{y2} \\
 \\
 y1\overline{r1} & y1\overline{g1} & y1\overline{y1} &
 y1\overline{r2} & y1\overline{g2} & y1\overline{y2} \\
\\
 r2\overline{r1} & r2\overline{g1} & r2\overline{y1} &
 r2\overline{r2} & r2\overline{g2} & r2\overline{y2} \\ \\
 g2\overline{r1} & g2\overline{g1} & g2\overline{y1} &
 g2\overline{r2} & g2\overline{g2} & g2\overline{y2} \\
 \\
 y2\overline{r1} & y2\overline{g1} & y2\overline{y1} &
 y2\overline{r2} & y2\overline{g2} & y2\overline{y2}
 \end{array}
\right).
\end{equation}
Four color singlet operators acting on the column vector $(r1, g1, y1, r2, g2, y2)^T$ are
\begin{equation}
\left(
 \begin{array}{cc}
  {\bf 0} &  {\bf 1}  \\
  {\bf 1} &  {\bf 0}
  \end{array}
\right),\ \
\left(
 \begin{array}{cc}
 {\bf 1} & {\bf 0}  \\
  {\bf 0} & -{\bf 1}
  \end{array}
\right),\ \
\left(
 \begin{array}{cc}
  {\bf 0} &  -i{\bf 1}   \\
 i{\bf 1} &  {\bf 0}
  \end{array}
\right),\label{SU2}
\end{equation}
and
\begin{equation}
\left(
 \begin{array}{cc}
 {\bf 1} & {\bf 0}   \\
 {\bf 0} & {\bf 1}
  \end{array}
\right)\  ,\label{U1An}
\end{equation}
where {\bf 1} and {\bf 0} are the $3\times 3$ identity and null matrices, respectively.
The generators of Eq. (\ref{SU2}) do not lead to a color anomaly from triangle diagrams, and the generator of Eq. (\ref{U1An}) is anomalous. It is required to give masses to Goldstone bosons corresponding to the generators of Eq. (\ref{SU2}), leaving the Goldstone boson corresponding to Eq. (\ref{U1An}) massless, except from the anomaly contribution, at this level. The Goldstone boson corresponding to Eq. (\ref{U1An}) is interpreted as the composite QCD axion $\ac$.

For this purpose, one may introduce two triplet flavon superfields $S^i_j (i,j=1,2)$ and $S'^i_j (i,j=1,2)$ with Tr$S^i_j=0$ and Tr$S'^i_j=0$, and the following superpotential
\begin{equation}
W=-gS^i_j \bar q_h^j \bar q_{hi}-g'S'^i_j \bar q_h^j \bar q_{hi}.
\end{equation}
Assigning VEVs to $\langle S^1_2\rangle=V$ and $\langle S'^1_1\rangle=-\langle S'^2_2\rangle=V'$, which are supposed to be much larger than the electroweak scale,  the flavor $SU(2)_f$ is completely broken and the Goldstone bosons corresponding to (\ref{SU2}) obtain mass. At this level, however, the $U(1)$ Goldstone boson corresponding to Eq. (\ref{U1An}) does not obtain mass.

Another method to make the $SU(2)_f$ triplet Goldstone bosons heavy is to make the flavor group $SU(2)_f$ be a gauged $SU(2)$. Then, $SU(2)_f$ gauge interactions make the $SU(2)_f$ triplet Goldstone bosons heavy as the colored Goldstone bosons become heavy by the QCD interactions of Fig. \ref{fig:coloredGoldstone}. Still at this level, the $U(1)$ Goldstone boson corresponding to Eq. (\ref{U1An}) remains massless. The $U(1)$ Goldstone boson obtains mass by the QCD anomaly, and becomes a composite QCD axion $\ac$.

Let us now proceed to discuss the two axion system of the composite axion $a_{\rm comp}$ and the MI axion (or the quintessential axion $a_{\rm quint}$).
We can write the instanton potential for $\aq=f\theta_q, \ac=UN_{DW}\theta$ (where $N_{DW}=nN_h$ with $n$ copies of  (\ref{hquarks}) is the domain wall number of the composite QCD axion \cite{Choi:1985iv}) and $\tilde\eta'_h=F_{\tilde\eta'_h}\theta_{\tilde\eta'_h}$ as
\begin{eqnarray}
&-\Lambda^4\cos(\theta_q+nN_h\theta)
-\lambda^4\cos(\theta_q+\alpha\theta+N_h
\theta_{\tilde\eta'_h}) \nonumber\\
&-m_{\tilde G_h}V^3\cos(\theta_{\tilde\eta'_h})
\label{axquintint}
\end{eqnarray}
where
$$\Lambda^4=\frac{Z}{(1+Z)^2}f_\pi^2 m_\pi^2,$$
and $V=\textrm{h-gluino condensation scale}.$
The MI axion couples to QCD and hidden sector with the same coupling, but the composite axion coupling to the hidden sector anomaly involves $\alpha$.  For (\ref{hquarks}), we have $\alpha=-3n$ due to three colors of QCD.  Integrating out the heavy $\tilde\eta'_h$, we set $\tilde\eta'_h=0$ in Eq. (\ref{axquintint}), and consider
\begin{equation}
-\Lambda^4\cos\left(\frac{\aq}{f}+\frac{\ac}{U}\right)
-\lambda^4\cos\left(\frac{\aq}{f}+\frac{\alpha\ac}{nN_hU}
\right)
\label{aqacpot}
\end{equation}
Then, for the composite QCD axion and the quintessential axion parametrized in (\ref{axquintint}), the following mass eigenvalues and eigenvectors in the $(\ac, \aq)$ basis are obtained in the limit $f\gg U$,
\begin{equation}
\mceq^2\simeq\frac{\Lambda^4+\lambda^4}{U^2},\ \ac\simeq \frac{1}{\sqrt{1+\delta^2}}\left(
\begin{array}{c}
 1\\ \delta
\end{array}
\right)
\end{equation}
\begin{equation}
\mqeq^2\simeq\frac{(1-\tilde\alpha)^2
\Lambda^4\lambda^4}{f^2(\Lambda^4+\lambda^4)},\ \aq\simeq \frac{1}{\sqrt{1+\delta^2}}\left(
\begin{array}{c}
 -\delta\\ 1
\end{array}
\right)
\end{equation}
where $\tilde\alpha=\alpha/nN_h$ and
$$
\delta\simeq \frac{U}{f}\left(\frac{\Lambda^4+\lambda^4}{ \Lambda^4+\tilde\alpha\lambda^4}\right) \left[1-(1-\tilde\alpha)^2\frac{\Lambda^4\lambda^4}{ (\Lambda^4+\tilde\alpha\lambda^4)^2}\right] .
$$
Thus, $a_{\rm comp}$ becomes the QCD axion and $\aq=a_{MI}$ becomes the quintessential axion. Now, $U$ is interpreted as the QCD axion decay constant $F_a$, and $f$ is interpreted as the quintessential axion decay constant \fq. Making the MI axion decay constant $f$ large, the quintessential axion (MI axion) can be made to account for DE. On the other hand, the QCD axion decay constant is related to the hidden sector confining scale which is in the axion window if it is detectable. As expected, $\alpha=N$ makes the quintessential axion the exact Goldstone boson since there is only one phase depending on two potential terms of Eq. (\ref{axquintint}).

\section{Discussion}
We proposed a field theoretic scenario for a quintessential axion from
a string theory motivation. The MI axion becomes a
quintessential axion while the confining force at an intermediate
scale generates the QCD axion scale through composite pseudo-scalar
mesons of h-squark condensation. The observable QCD axion
as well as the DE problem have been important issues in string theory.
The observable QCD axion from superstring could come from
an approximate global symmetry \cite{CKKapprox,Choi:2009jt}.
Here, we considered the composite axion idea which has not been
discussed before in that context.
The mechanism depends strongly on the fact that the hidden
sector anomaly term could be much smaller than the QCD anomaly term. This
feature was anticipated before \cite{KimNilles03}, and here we
present an explicit estimate of h-quark mass as well as the mass of
the composite QCD axion.
The small value of the quintessential axion mass (and the value of DE)
is based on the fact that there are
many light h-quarks below the scale $\Lambda_h$, in contrast to
the situation in QCD with only three light quarks below
$\Lambda_{\rm QCD}$. Simply, by counting engineering dimensions
with a sufficient number of h-quarks one notes that the hidden
sector instanton potential can be made negligible compared to the
QCD anomaly term. In heterotic orbifold compactification, our idea
is not necessarily realized with a small number of h-quarks as
noted in the very promising flipped SU(5) and MSSM
models \cite{heterotichidden}. So, even though we presented a
field theoretic study of the composite QCD axion here, it is a
difficult problem to find such models easily in from string
compactification.
But the landscape study reveals many models with enough
h-quarks \cite{landscape} that have to be checked with SUSY
phenomenology.
A further suppression of the DE-scale could be abtained
along the lines as suggested in
\cite{Lowen:2008xh}. In the future, we might extend our study to include
hidden sectors with chiral h-quarks \cite{KimGMSBst}.

\vskip 0.3cm
\acknowledgments{
 J.E.K. is supported in part by the Korea Research Foundation,
Grant No. KRF-2005-084-C00001, and H.P.N. is supported in part
by the
European Union 6th framework program  MRTN-CT-2006-035863 ``UniverseNet''
and SFB-Transregio 33 "The Dark Universe" by Deutsche
Forschungsgemeinschaft (DFG). .
}

\end{document}